\documentclass[
 reprint,
 amsmath,amssymb,
 aps,
 prl,
]{revtex4-2}

\usepackage{graphicx}
\usepackage{xcolor}
\usepackage{dcolumn}
\usepackage{bm}

\usepackage{amsmath,amssymb,amsthm,mathtools,amsfonts,mathrsfs}
\usepackage{algorithm}
\usepackage{algpseudocode}
\usepackage[version=4]{mhchem}

\usepackage{hyperref}
\usepackage{url}

\begin{document}

\title{Asymptotics of Protein Number Distribution in Stochastic Gene Expression Models under Burst Approximation}

\author{Yuntao Lu}
\email{yuntaolu22@m.fudan.edu.cn}

\author{Yunxin Zhang}
\email{xyz@fudan.edu.cn}

\affiliation{
School of Mathematical Sciences, Fudan University, Shanghai 200433, CHINA
}

\date{\today}

\begin{abstract}
The burst approximation is a widely used technique to simplify stochastic gene expression models. However, the dynamics and analytical properties of the protein number distribution in gene expression models under the burst approximation are barely studied. In this study, we propose and systematically analyze surrogate models with multiple gene states and arbitrary burst size distributions. An analytical time-dependent solution to the chemical master equation is derived and then exploited in two directions. Theoretically, several fine properties of the protein number distribution are established using functional analysis. For geometrically distributed burst sizes, the distribution is dominated by a scaled negative binomial distribution, and is light-tailed in certain parameter regimes. Computationally, we develop efficient algorithms in three settings, enabling fast calculation of the protein number distribution. Furthermore, the approximation error relative to full gene expression models is estimated in terms of low-order moments of the distribution, thereby clarifying the validity of the burst approximation.
\end{abstract}
\maketitle

Gene expression is inherently stochastic, as transcription, translation, and molecular degradation are all driven by random molecular events. This intrinsic stochasticity becomes particularly pronounced in cells where relevant biomolecules exist in low copy numbers \cite{CentralDogmaSinglemolecule2011}. Consequently, even genetically identical cells in the same environment may exhibit substantial cell-to-cell variability in messenger RNA (mRNA) and protein abundance \cite{StochasticMRNASynthesis2006}. Such variability has been directly observed in single-cell and single-molecule experiments, including real-time measurements of transcriptional bursting in bacteria \cite{RealTimeKineticsGene2005}, and single-molecule studies of protein production in living cells \cite{ProbingGeneExpression2006,StochasticProteinExpression2006}. The molecular mechanism underlying the transcriptional bursting has also been investigated \cite{TheoreticalInvestigationTranscriptional2018,MechanochemicalModelTranscriptional2020,UnderstandingMolecularMechanisms2021}.

\begin{figure}[h]
    \centering
    \includegraphics[width=0.8\linewidth]{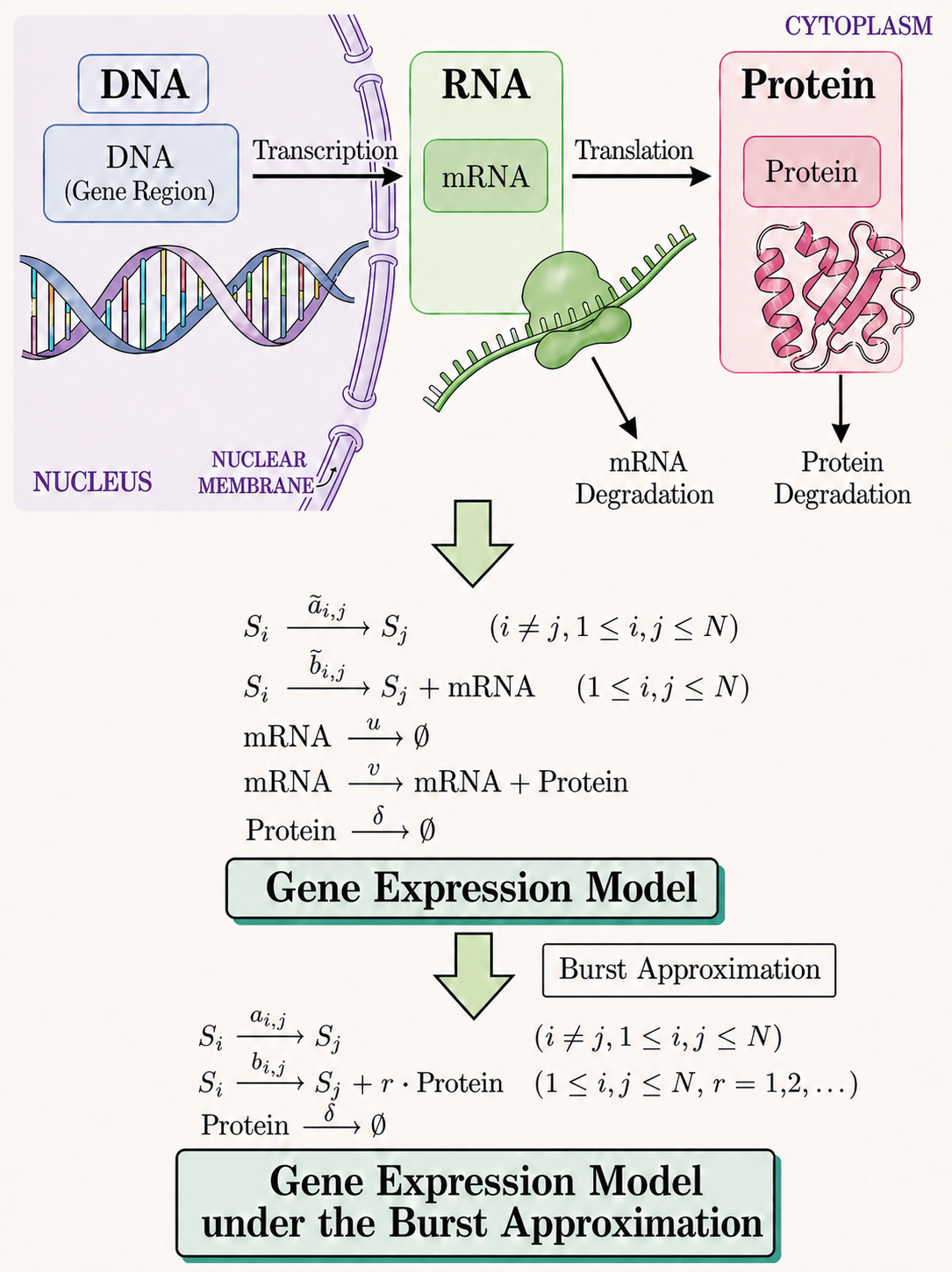}
    \caption{(\textbf{Workflow of Gene Expression Modeling}) In a typical gene expression process, transcription in the nucleus produces mRNA, which is then translated into protein in the cytoplasm. Both mRNA and protein are subject to degradation. The full gene expression model describes switching among multiple gene states, with productive transitions generating mRNA, followed by mRNA degradation, translation, and protein degradation. Under the burst approximation, the explicit mRNA intermediate is eliminated and replaced by protein bursts.}
    \label{fig:workflow}
\end{figure}

The continuous-time Markov chain is a standard tool for describing the interaction of multiple components in a stochastic chemical reaction system, and its Kolmogorov forward equation is conventionally termed the chemical master equation (CME) in such context \cite{StochasticProcessesPhysics2007,StochasticMethodsHandbook2009,StochasticAnalysisBiochemical2015,ApproximationInferenceMethods2017,AppliedStochasticAnalysis2019,StochasticChemicalReaction2021}. Gene expression can be readily modeled using the CME by assuming that transcription, translation, and molecular degradation are all first-order reactions, as shown in \autoref{fig:workflow}. However, for full gene expression models, the probability distribution of the protein copy number is analytically intractable even when the gene remains active \cite{AnalyticalDistributionsStochastic2008}. The binomial moment method \cite{BinomialMomentEquations2011,StochasticAnalysisComplex2012,MomentconvergenceMethodStochastic2016} is applied to a model with two gene states (one active, one inactive) in \cite{ExactDistributionsFull2017}, where both the recurrence relation for binomial moments and analytical expressions for low-order moments are provided. Similar methodology is also used to study a model with multiple gene states \cite{InfluenceComplexPromoter2019}. 
Given the difficulty of analyzing full gene expression models, the burst approximation is conceptually proposed as an approximation technique to simplify the full models, building upon the experimental conclusion that mRNAs decay substantially faster than proteins in most cells \cite{AnalyticalDistributionsStochastic2008}. Remarkably, gene expression models under the burst approximation admit fruitful analytical results. Two models where the gene remains active and switches between two states (one active, one inactive) are studied in \cite{AnalyticalDistributionsStochastic2008} using the generating function method, where the analytical probability mass function (PMF) of the protein copy number is derived in steady state. Recently, the binomial moment method has been applied to models with multiple gene states \cite{ExactDistributionsStochastic2022}, where a recurrence relation among binomial moments is formally obtained. The analytical PMF is presented for models with one active gene state. Through mapping models to $GI^X/G/\infty$ queueing systems \cite{GIXInftySystem1990,SolvingStochasticGeneexpression2024}, non-Markovian models can also be studied using queueing theory \cite{IntrinsicNoiseStochastic2011,TranscriptionalBurstingGene2015}. The correspondence between models under the burst approximation and queueing systems with batch arrivals may explain the substantial analytical simplifications introduced by the burst approximation. Additionally, in certain non-Markovian models, the steady-state distribution of the protein copy number is identical to that of constructed Markovian counterparts \cite{MarkovianApproachesModeling2019,AnalyticalResultsNonMarkovian2020}, which enables steady-state analysis via the corresponding Markovian models.

However, on the one hand, previous studies mainly focus on numerical computation of the protein number distribution, therefore few general properties of the distribution are known, except for low-order moments. On the other hand, existing works mostly consider the steady-state distribution of the protein copy number, whereas the dynamics is rarely studied. Additionally, whether errors introduced by the burst approximation are controllable has not received much attention. The validity of the burst approximation has been assessed only in the model where the gene remains active \cite{AnalyticalDistributionsStochastic2008}.

We propose a general gene expression model under the burst approximation, represented as a chemical reaction system \eqref{Reaction}. In this model, the gene has $N$ distinct metastable states, namely, $S_i\;(1\leq i \leq N)$, and it switches arbitrarily among these states memorylessly. Within each metastable state, proteins are produced in bursts at state-dependent rates, and the gene state may be altered during protein production. Note that during one burst, multiple protein molecules can be generated simultaneously, and the exact number follows a probability distribution. Existing protein molecules degrade at a given rate independent of the gene state and the protein count. Protein autoregulation of gene expression is not considered \cite{SteadystateFluctuationsGenetic2012}. Following the stochastic description of the chemical reaction kinetics \cite{AppliedStochasticAnalysis2019}, the model \eqref{Reaction} can be mathematically interpreted as a continuous-time Markov chain, whose dynamics is governed by the CME. 

To be specific, let $S(t)$, $M(t)$ denote the gene state and the protein count in the system at time $t$.
Let $a_{i,j}\;(i\neq j,\;1\leq i,j\leq N)$ denote the transition rates between gene states without production of proteins; $b_{i,j}^{[r]}\;(1\leq i,j\leq N)$ denote the transition rates of generating $r$ protein molecules in a single burst and the gene transitioning from $S_i$ to $S_j$; $\delta$ denote the degradation rate of proteins. Define $a_{i,i}:=-\sum_{\substack{k=1\\k\neq i}}^Na_{i,k}-\sum_{k=1}^N\sum_{r=1}^\infty b_{i,k}^{[r]}$ for $1\leq i\leq N$, $D_0:=(a_{i,j})_{N\times N}$, $D_r:=(b_{i,j}^{[r]})_{N\times N}$ for $r=1,2,\cdots$, and $D:=D_0 + \sum_{r=1}^\infty D_r$. Assume throughout this study that the sequence $\{b^{[r]}_{i,j}\}_{r\geq1}$ has finite moments of all orders, namely $\sum_{r=1}^\infty r^nb^{[r]}_{i,j}<\infty,\;\forall\;n\in\mathbb{N},1\leq i,j\leq N$. In particular, $\sum_{r=1}^\infty b^{[r]}_{i,j}$ is finite for $1\leq i,j\leq N$, hence $D$ is well-defined. For simplicity, we refer to $N$ as the order of the model \eqref{Reaction}.

\begin{equation}\label{Reaction}
\begin{aligned}
    &\ce{S_i ->[$a_{i,j}$] S_j}\;(i\neq j,\;1\leq i,j\leq N)\\
    &\ce{S_i ->[$b_{i,j}^{[r]}$] S_j + r \cdot Protein}\;(1\leq i,j\leq N,\;r\geq 1)\\
    &\ce{Protein ->[$\delta$] $\emptyset$}\\
\end{aligned}
\end{equation}

Let $\mathbb{P}_{i,j}(m;t)$ denote the probability that $S(t)=S_j$ and $M(t)=m$, given the Dirac initial distribution concentrated at $S(0)=S_i$ and $M(0)=0$. The CME governing $\mathbb{P}_{i,j}(m;t)$ is ($1\leq i,j\leq N,m\in\mathbb{N}$)
\begin{equation}\label{CME}
\begin{aligned}
\frac{\partial}{\partial t}&\mathbb{P}_{i,j}(m;t)
=
\sum_{s=1}^N a_{s,j}\mathbb{P}_{i,s}(m;t)\\&
+
\sum_{r=1}^{\infty}\sum_{s=1}^N b_{s,j}^{[r]}
\mathbb{P}_{i,s}(m-r;t)\\&
+(m+1)\delta\,\mathbb{P}_{i,j}(m+1;t)
-m\delta\,\mathbb{P}_{i,j}(m;t).
\end{aligned}
\end{equation}
where $\mathbb{P}_{i,j}(m;t)$ is conventionally taken as zero when $m<0$. 
Using the generating function method, we conclude that the analytical solution to \eqref{CME} is, for $m=0$, 
\begin{equation}\label{sol0}
\begin{aligned}
    \mathbb{P}_{i,j}(0;t)=&\bm{e}_i^\top\mathrm{e}^{D_0t}\bm{e}_j\\&+\sum_{k=1}^\infty\sum _{l_1,\cdots,l_k\geq1}\int_{\Omega_k}\Bigg[\prod_{s=1}^{k}\bigl(1-\alpha(t,t_s)\bigr)^{l_s}\Bigg]\\&\cdot\bm{e}_i^\top
K(t;t_1,\cdots,t_k;l_1,\cdots,l_k)\bm{e}_j\mathrm{d}\bm{t},
\end{aligned}
\end{equation}
and, for $m\geq1$,
\begin{equation}\label{sol}
\begin{aligned}
\mathbb P_{i,j}(m;t)=&\sum_{k=1}^{\infty}\sum _{l_1,\cdots,l_k\geq1}\int_{\Omega_k}\Bigg[\sum_{\substack{m_1+\cdots+m_k=m\\0\leq m_1\leq l_1,\cdots,0\leq m_k\leq l_k}}\\&
\prod_{s=1}^k \binom{l_s}{m_s}\alpha(t,t_s)^{m_s}(1-\alpha(t,t_s))^{l_s-m_s}\Bigg]\\&\cdot
\bm{e}_i^\top K(t;t_1,\cdots,t_k;l_1,\cdots,l_k)\bm{e}_j\mathrm{d}\bm{t}.
\end{aligned}
\end{equation}
In \eqref{sol0} and \eqref{sol}, $\Omega_k:=\{(t_1,t_2,\cdots,t_k)\mid0\leq t_1\leq t_2\leq\cdots\leq t_k\leq t\}\subseteq\mathbb{R}^k$; $\alpha(t,t_s):=\exp(-\delta\cdot t+\delta\cdot t_s)$; $\binom{l_s}{m_s}$ denotes the binomial coefficient; and $\{\bm{e}_s\}_{s=1}^N$ are standard basis vectors in $\mathbb{R}^N$ where $\bm{e}_s$ is the column vector with a $1$ in the $s$-th position and $0$ elsewhere. The family of functions $\{K(t;t_1,\cdots,t_k;l_1,\cdots,l_k)\}_{k\geq 1}$ is defined as follows: $K(t;t_1;l_1):=\mathrm{e}^{D_0t_1}D_{l_1}\mathrm{e}^{D_0(t-t_1)}$, $K(t;t_1,t_2;l_1,l_2):=\mathrm{e}^{D_0t_1}D_{l_1}\mathrm{e}^{D_0(t_2-t_1)} D_{l_2}\allowbreak
\mathrm{e}^{D_0(t-t_2)}$, and
$K(t;t_1,\cdots,t_k;l_1,\cdots,l_k):=\mathrm{e}^{D_0t_1}D_{l_1}\allowbreak\mathrm{e}^{D_0(t_2-t_1)}\allowbreak D_{l_2}\cdots D_{l_{k-1}}\mathrm{e}^{D_0(t_k-t_{k-1})} D_{l_{k}}\allowbreak
\mathrm{e}^{D_0(t-t_k)}$ for $k\geq3$.
The exponential of a square matrix, say $C$, is defined as $\mathrm{e}^C:=\sum_{k=0}^\infty C^k/k!$. 

Although \eqref{sol0} and \eqref{sol} are not directly suitable for numerical computation, they serve as a starting point for further analysis, which yields fine properties of the steady-state probability distribution.
We first proceed by calculating the time-dependent expressions for binomial moments of the protein number distribution, and take the temporal limit. This is partially motivated by the well-known binomial moment method. The (matrix-form) binomial moments are defined as $[\mathcal{B}_m(t)]_{i,j}:=\sum_{n=m}^\infty\binom{n}{m}\mathbb{P}_{i,j}(n;t),\;1\leq i,j\leq N,\;m\in\mathbb{N}$. For technical reasons, we mainly consider the coarse-grained (scalar) binomial moments (in steady state), namely $B_m:=\lim_{t\rightarrow\infty}\bm{\pi}^\top\mathcal{B}_m(t)\bm{1},\;m\in\mathbb{N}$, where $\bm{\pi}\in\mathbb{R}^{N\times1}$ is the invariant distribution of the underlying Markov chain characterized by $D$, and $\bm{1}\in\mathbb{R}^{N\times1}$ is a column vector with all ones. Throughout the study, we assume that $D$ is irreducible, so that it admits a unique invariant distribution. The coarse-grained (scalar) PMF (in steady state) is defined similarly by $P_n:=\lim_{t\rightarrow\infty}\bm{\pi}^\top\mathbb{P}(n;t)\bm{1},\;n\in\mathbb{N}$, where $[\mathbb{P}(n;t)]_{i,j}:=\mathbb{P}_{i,j}(n;t),\;1\leq i,j\leq N,\;n\in\mathbb{N}$.

We conclude that binomial moments in steady state are all finite and have the following expressions:
\begin{equation}\label{binomial}
\begin{aligned}
B_1
&= \frac{1}{\delta}\bm{\pi}^{\top} C_1 \bm{1},
\\[1mm]
B_m
&= \frac{1}{m\delta}
\sum_{k=1}^m
\sum_{\substack{
l_1+\cdots+l_k=m\\
l_1,\ldots,l_k\geq 1
}}
\bm{\pi}^{\top} C_{l_1}
\left(l_1\delta\bm{I}_N-D\right)^{-1}
C_{l_2}
\\
&\quad \cdot
\left[(l_1+l_2)\delta\bm{I}_N-D\right]^{-1}
C_{l_3}
\cdots
\\
&\quad \cdot
\left[(l_1+\cdots+l_{k-1})\delta\bm{I}_N-D\right]^{-1}
C_{l_k}\bm{1},
\; m\geq 2 .
\end{aligned}
\end{equation}
where $C_r:=\sum_{n=r}^\infty\binom{n}{r}D_n$ and $\bm{I}_N$ denotes the $N$-dimensional identity matrix. Note that $\{C_r\}_{r\geq 1}$ are all well-defined since we assume moments of all orders of $\{b^{[r]}_{i,j}\}_{r\geq1}$ exist. In \eqref{binomial}, the summation is taken over all ordered partitions of $m$ into positive integers. For example, when $m=3$, the summation includes four ordered partitions: $l_1=3$; $l_1=1,l_2=2$; $l_1=2,l_2=1$; and $l_1=1,l_2=1,l_3=1$. From \eqref{binomial}, one can readily calculate low-order binomial moments and, consequently, important summary statistics such as the noise and the Fano factor. The analytical results are reported in the Supplemental Materials and are consistent with existing studies \cite{ExactDistributionsStochastic2022}.
In general, direct computation based on \eqref{binomial} quickly becomes impractical as $m$ grows, because of the combinatorial enumeration needed to determine the integer partition of $m$. Actually, matrix computations within the summation is relatively fast, therefore the computational bottleneck is independent of the order of the model ($N$) and is purely combinatorial. We will also provide a combinatorial trick to overcome it in certain cases.

Note that, if $D_r=c_rD_1\;(r\geq2)$ for some constants $c_r\geq0\;(r\geq2)$, and $D_1$ is diagonal, the model \eqref{Reaction} reduces to the one in \cite{ExactDistributionsStochastic2022}.

We derive the upper bound for the PMF by repeatedly applying an analogue to Theorem 4.1.2 in \cite{MatrixComputations2013}, the H{\"o}lder inequality, and the submultiplicativity of operator norm. To be specific, we have (for $n\geq 1$)
\begin{equation}\label{upper bound}
\begin{aligned}
P_n \leq
&\sum_{k=1}^n
  \sum_{\substack{
    l_1+\cdots+l_k=n\\
    l_1,\ldots,l_k\geq 1
  }}
\frac{
  \prod_{i=1}^k \lVert C_{l_i}\rVert_{\infty}
}{
  \delta^k
  l_1(l_1+l_2)\cdots(l_1+\cdots+l_k)
}.
\end{aligned}
\end{equation}
where the infinity norm of a matrix, denoted by $\lVert \cdot\rVert_\infty$, is the maximum absolute row sum of this matrix.

In general, \eqref{upper bound} cannot be further simplified. However, when $\{D_r\}_{r\geq 1}$ follows a geometric distribution, we obtain elegant results. We note that this assumption is reasonable, since the number of proteins generated during one burst has been experimentally shown to approximately follow a geometric distribution \cite{StochasticProteinExpression2006,ProbingGeneExpression2006}.
To be specific, we now assume $D_r=\lambda^{r-1}D_1\;(r\geq1)$, where $\lambda\in(0,1)$ is the parameter of a geometric distribution. \eqref{upper bound} then reduces to:
\begin{equation}\label{upper bound2}
\begin{aligned}
P_n\leq\frac{1}{n!}\left(\frac{\lambda}{1-\lambda}\right)^n\left[\frac{\lVert D_{1}\rVert_{\infty}}{\delta\lambda(1-\lambda)}\right]^{(n)},\;n\in\mathbb{N}.
\end{aligned}
\end{equation}
For any real number $\zeta$, $\zeta^{(m)}:=\Gamma(\zeta+m)/\Gamma(\zeta)$ is the Pochhammer symbol, where $\Gamma(\zeta)$ is the Gamma function. A combinatorial identity is developed to prove \eqref{upper bound2}, which overcomes the complexity bottleneck of enumerating ordered partitions of integers.
The upper bound in \eqref{upper bound2} converges to zero if and only if $\lambda\in(0,1/2)$ or $\lambda=1/2$ and $\lVert D_{1}\rVert_{\infty}/\delta<1/4$.
It follows from \eqref{upper bound2} that the PMF of the protein copy number is bounded from above by a constant multiple of a negative binomial distribution.
See Figure S.1 for numerical verification.

When $\lambda\in(0,1/2)$, the steady-state probability distribution of the protein copy number is light tailed, since the moment generating function $\mathcal{M}[\{P_n\}](s):=\sum_{n=0}^\infty P_n\exp(ns)$ is finite for $s\in(0,\ln(1-\lambda)-\ln\lambda)$. Based on the Chernoff method \cite{ConcentrationInequalitiesNonasymptotic2013}, we also estimate the tail probability of protein copy number. To be specific, for $m\in\mathbb{N}$, we have
\begin{equation}\label{tail}
\begin{aligned}
\sum_{n=m}^{\infty} P_n
\leq&
\left[
\frac{\lVert D_1\rVert_\infty
      +m\delta\lambda(1-\lambda)}
     {m\delta(1-\lambda)^2}
\right]^m
\\&\quad\cdot
\left[
1+
\frac{m\delta\lambda(1-\lambda)}
     {\lVert D_1\rVert_\infty}
\right]^{
\frac{\lVert D_1\rVert_\infty}
     {\delta\lambda(1-\lambda)}
}.
\end{aligned}
\end{equation}
The above estimation of the moment generating function and the tail probability generalizes the Heavy-Tailed Law for transcription models \cite{ExtrinsicNoiseHeavyTailed2020,ExactlySolvableModels2020,StochasticKineticsMRNA2025}.  

There are two motivations for examining the recurrence relation among subsequent binomial moments. First, according to \eqref{binomial}, many terms in the summation are repeatedly recomputed if all binomial moments are computed independently. Second, the binomial moment equations \cite{BinomialMomentEquations2011} are decoupled for the first-order chemical reaction system \eqref{Reaction}, therefore higher-order binomial moments are linear combinations of lower-order binomial moments in steady state. In this study, we derive the recurrence relation directly from \eqref{binomial}. A formal but more direct approach is to apply the binomial moment method as performed in \cite{ExactDistributionsStochastic2022}. 
Based on \eqref{binomial}, we have
\begin{equation}\label{recurrence}
\begin{aligned}
\mathcal{B}_0&=\bm{I}_N,\;\;
\mathcal{B}_m:=\left(\sum_{n=1}^m\mathcal{B}_{m-n}C_n\right)\left(m\delta\bm{I}_N-D\right)^{-1},\\
B_0&=1,\;\;B_m=\frac{1}{m\delta}\bm{\pi}^\top\left(\sum_{n=1}^m\mathcal{B}_{m-n}C_n\right)\bm{1},\;m\geq 1.
\end{aligned}
\end{equation}
A numerical implementation of \eqref{recurrence} is given in Algorithm S.1.

According to Algorithm S.1, binomial moments up to arbitrary orders can be obtained recursively. Formally, the PMF can then be constructed based on the following identity \cite{MomentconvergenceMethodStochastic2016}:
\begin{equation}\label{pmf}
   P_n=\sum^{\infty}_{m=n}\left(-1\right)^{m-n}\binom{m}{n}B_m,\; n\in\mathbb{N}.
\end{equation}
However, for the series \eqref{pmf} to converge, the term $\binom{m}{n}B_m$ should converge to zero as $m$ grows, implying that the sequence $\{B_m\}_{m\in\mathbb{N}}$ converges to zero. We prove that when the burst size follows a geometric distribution with parameter $\lambda\in(0,1/2)$, the PMF can be reconstructed through \eqref{pmf}. When the burst size follows a geometric distribution with parameter $\lambda\in[1/2,1)$, the series \eqref{pmf} may diverge and the PMF cannot be readily recovered. In Figure S.2, we provide examples where all orders of binomial moments exist but $\lim_{m\rightarrow\infty}B_m=\infty$.

In addition, truncation error may be introduced during the computation of $\{C_r\}_{r\geq 1}$. Fortunately, analytical expressions for $\{C_r\}_{r\geq 1}$ can be derived under three distinct yet reasonable assumptions.

\textbf{(a) Burst Size Follows Geometric Distribution}. This assumption has been introduced above. When $D_r=\lambda^{r-1}D_1\;(r\geq1)$ for some given $\lambda\in(0,1)$, we have (for $r\geq 1$)
\begin{equation}\label{geo}
\begin{aligned}
    &C_r=\frac{\lambda^{r-1}}{(1-\lambda)^{r+1}}D_1,\;D=D_0+\frac{1}{1-\lambda}D_1.
\end{aligned}
\end{equation}
Across several examples, we compare our results with those obtained from the stochastic simulation algorithm \cite{GeneralMethodNumerically1976,ExactStochasticSimulation1977,StochasticSimulationChemical2007} and the finite state projection algorithm \cite{FiniteStateProjection2006}, as shown in Figure S.3.

\textbf{(b) Burst Size Follows Poisson Distribution}. When the burst size follows a Poisson distribution, namely, $D_r=(\alpha^{r-1}/r!)D_1,\;r\geq1$ for some positive real number $\alpha$, we also obtain analytical expressions (for $r\geq 1$):
\begin{equation}\label{poi}
\begin{aligned}
&C_r=\frac{\alpha^{r-1}}{r!}\mathrm{e}^\alpha D_1,\;D=D_0+\frac{1}{\alpha}(\mathrm{e}^\alpha-1)D_1.
\end{aligned}
\end{equation}

\begin{figure}[h]
    \centering
    \includegraphics[width=0.7\linewidth]{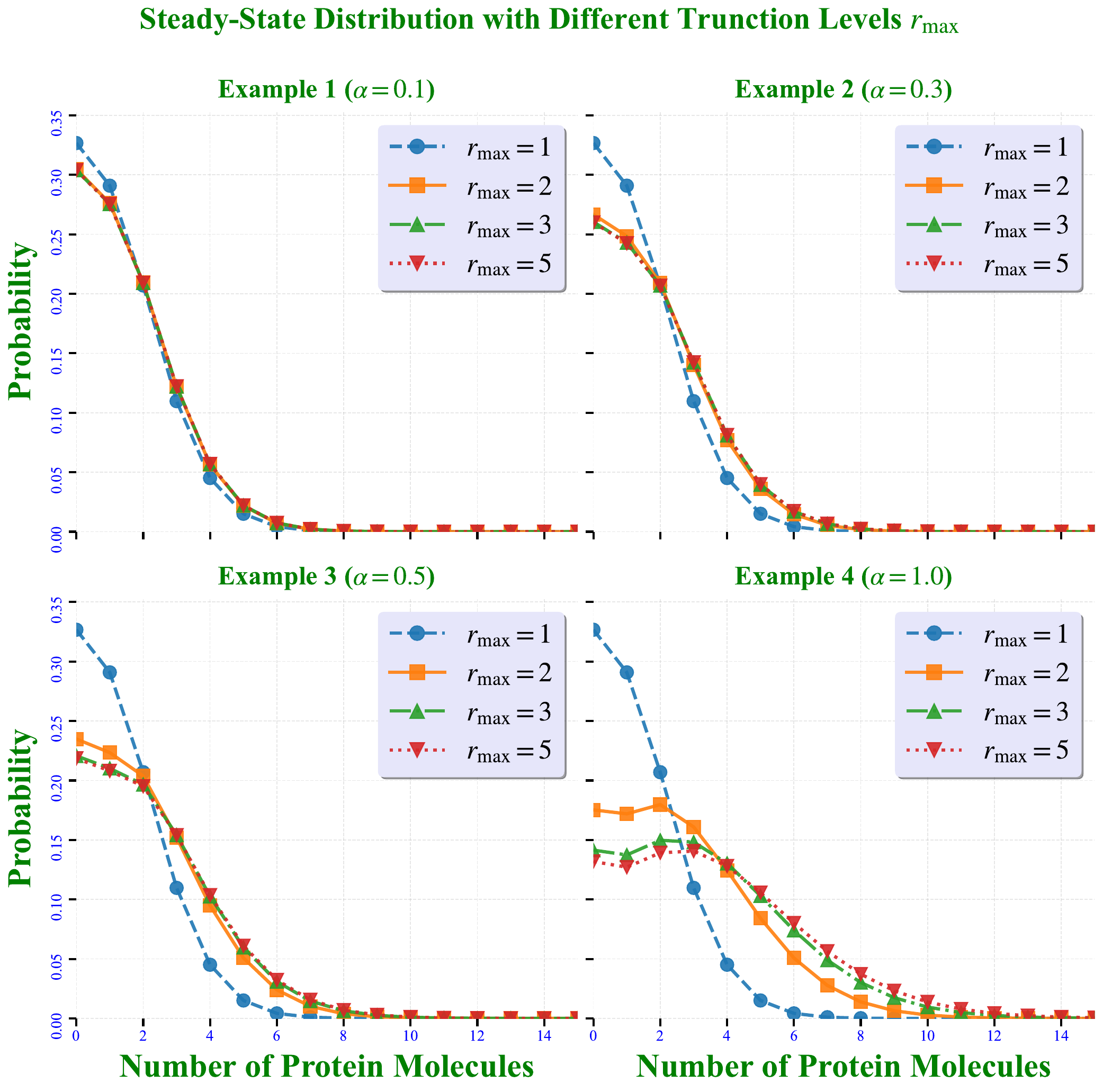}
    \caption{(\textbf{Probability Distribution of Protein Copy Number with Different Truncation Levels}) The burst size is assumed to follow a Poisson distribution parameterized by $\alpha$. The figure shows the steady-state PMF of the protein number with varying truncation levels $r_{\max}$. The four panels correspond to $\alpha=0.1$, $0.3$, $0.5$, and $1.0$, and within each panel the distributions obtained with $R_{\max}=1,2,3,$ and $5$ are compared. $D_0=\bm{0}_{4\times 4}$, $D_1=\big((0.2,0.2,0.1,0.8),(0.2,0.2,0.1,0.8),(0.1,0.1,0.1,0),(0.8,0.8,\allowbreak0,0.8)\big)$, $\delta=1$, and the burst size is geometrically distributed.}
    \label{PoissonTruncation}
\end{figure}

\textbf{(c) Burst Becomes Negligible for Large Sizes}. During the gene expression process, the probability that large numbers of proteins are simultaneously produced is negligible. Therefore, we may assume $D_r=\bm{0}_{N\times N}$ for $r$ strictly exceeding a given threshold, denoted as $r_{\mathrm{max}}$. In such cases, $C_r=\bm{0}_{N\times N}$ for $r\geq  r_{\mathrm{max}}+1$, and the series $C_r=\sum_{n=r}^{r_{\mathrm{max}}}\binom{n}{r}D_n,\;1\leq r\leq r_{\mathrm{max}}$ only involves finite terms. This assumption is also a prerequisite for numerically studying models in \eqref{CME} via the stochastic simulation algorithm and the finite state projection algorithm, since they only admit a finite number of reaction pathways. Empirically, as the truncation level $r_{\mathrm{max}}$ increases, the computed probability distribution converges relatively fast to the exact distribution corresponding to the full (non-truncated) model. See \autoref{PoissonTruncation} for examples.

We now connect our results to the classic work \cite{AnalyticalDistributionsStochastic2008} by studying the simplest case where $N=1$ in \eqref{Reaction} and the burst size of proteins is geometrically distributed. 
Assume $\{D_r\}_{r\geq 1}$ follows a geometric distribution with parameter $\lambda\in(0,1/2)$ and $N=1$. Note that $D_1$ is a positive real number.
Based on \eqref{binomial} and \eqref{pmf}, we derive that the steady-state distribution of the protein count is a negative binomial distribution (for $n\in\mathbb{N}$):
\begin{equation}\label{NBD}
\begin{aligned}
    P_n=\frac{\lambda^n}{n!}\left[\frac{D_1}{\delta \lambda(1-\lambda)}\right]^{(n)}(1-\lambda)^{D_1/[\delta \lambda(1-\lambda)]},
\end{aligned}
\end{equation}
which is consistent with existing results \cite{AnalyticalDistributionsStochastic2008}.

Finally, we compare low-order moments of the protein number distribution in full gene expression models and models under the burst approximation. Upper bounds of the difference between low-order moments are established using functional analysis. Recall that, as shown in \autoref{fig:workflow}, the burst approximation is conceptually an approximation technique to simplify the following full gene expression model:
\begin{equation}\label{Reaction1}
\begin{aligned}
    &\ce{S_i ->[$\widetilde{a}_{i,j}$] S_j}\;\;(i\neq j,\;\;1\leq i,j\leq N)\\
    &\ce{S_i ->[$\widetilde{b}_{i,j}$] S_j + mRNA}\;\;(1\leq i,j\leq N)\\
    &\ce{mRNA ->[$u$] $\emptyset$}\\
    &\ce{mRNA ->[$v$] mRNA + Protein}\\
    &\ce{Protein ->[$\delta$] $\emptyset$}\\
\end{aligned}
\end{equation}
In \eqref{Reaction1}, the parameter $\delta$ carries the same meaning as in \eqref{Reaction}, while $\widetilde{a}_{i,j},\;\widetilde{b}_{i,j}\;(1\leq i,j\leq N)$, $v$, and $u$ denote the transition rates among gene states without transcription, the transition rates with production of one mRNA molecule, the translation rate, and the mRNA degradation rate.

Under the notations in \eqref{Reaction} and \eqref{Reaction1}, two models are related by the following identities.
\begin{equation}\label{appr1}
\begin{aligned}
b_{i,j}^{[r]}&=\left(\frac{v}{u+v}\right)^r\frac{u}{u+v}\,\widetilde{b}_{i,j},\;1\leq i,j\leq N,\;r\geq1,\\
\widetilde{a}_{i,j}&=a_{i,j}-\frac{u}{u+v}\widetilde{b}_{i,j},\;1\leq i,j\leq N.
\end{aligned}
\end{equation}
The relation \eqref{appr1} can be interpreted as follows. Once transcribed, an mRNA molecule is subject to two competing reaction pathways, namely, degradation and translation. More specifically, this can be seen as the competing binding of decay complexes that promote degradation, and the recruitment of initiation factors that engage the ribosome for translation. Since the probability of initiating translation rather than degradation is $v/(u+v)$, the number of protein molecules produced from a single mRNA molecule follows a geometric distribution with parameter $u/(u+v)$. Hence, \eqref{appr1} readily follows.

\begin{figure}[h!]
    \centering
    \includegraphics[width=0.9\linewidth]{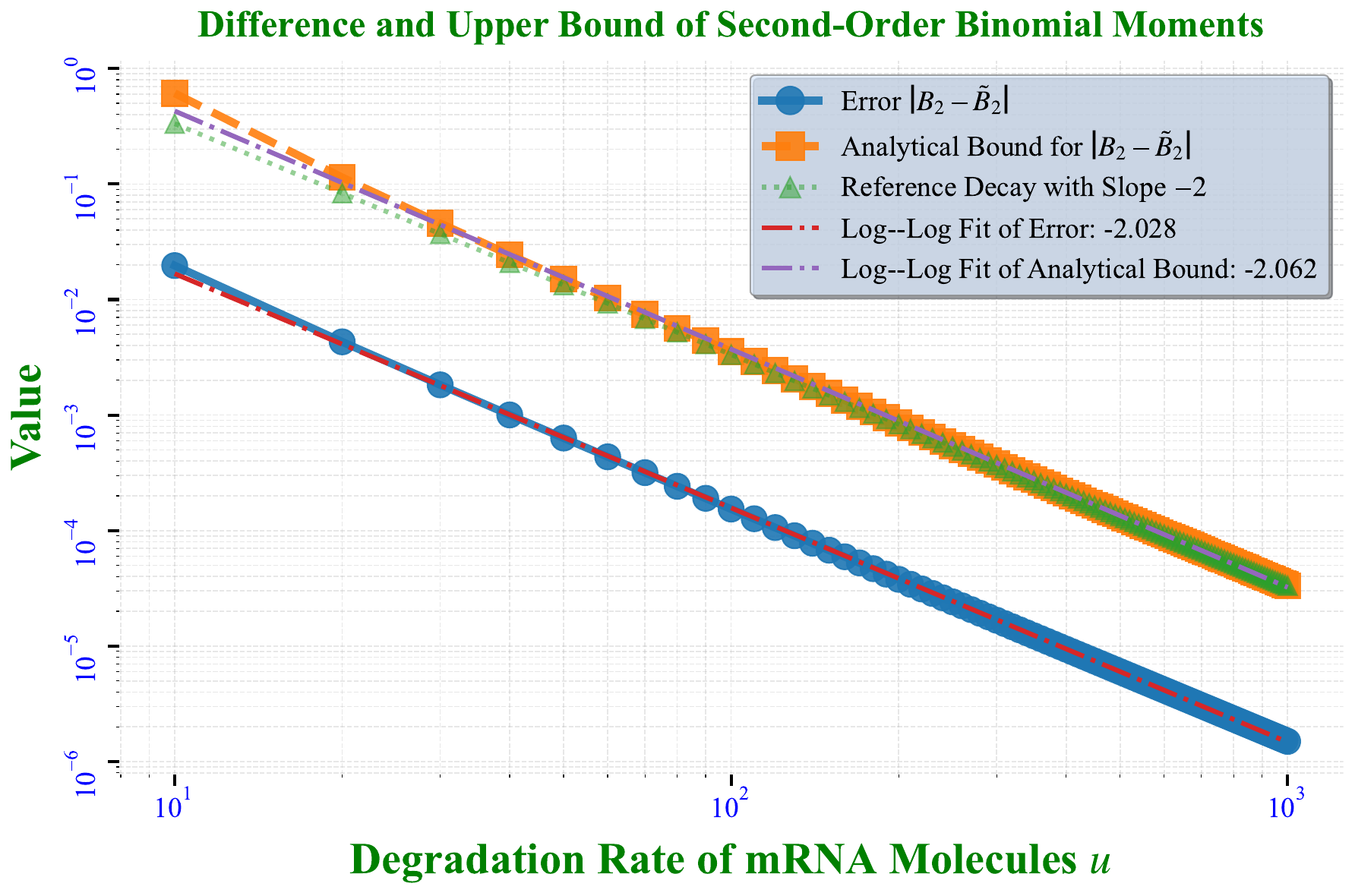}
    \caption{(\textbf{Approximation Error with respect to Second-Order Binomial Moments}): The absolute error $\lvert B_{2}-\widetilde{B}_{2}\rvert$ and its analytical upper bound \eqref{accuracy} are shown, plotted against the mRNA degradation rate $u$ on a logarithmic scale. Dashed lines with square markers indicate the bound, solid lines with circular markers indicate the computed absolute error, and the dotted line represents a reference decay rate. Dash-dotted lines show the corresponding log-log fitted slopes. In this example, $D_0$, $D_1$, and $\delta$ are the same as those in \autoref{PoissonTruncation}. Additionally, $\lambda=0.7$, $v=2$, and $u\in[10,1000]$.}
    \label{fig:error_bound}
\end{figure}

Low-order moments of mRNA and protein copy numbers in \eqref{Reaction1} can be formally obtained by applying the binomial moment method \cite{ExactDistributionsFull2017,InfluenceComplexPromoter2019}. Compared with \eqref{binomial}, we conclude that the first-order binomial moment, namely the expectation of the protein copy number, remains exact under the burst approximation; while higher-order binomial moments are explicitly altered when the burst approximation is applied. In particular, upper bounds can be derived for the absolute difference between low-order binomial moments from two models. For example, denoting the second-order binomial moment of the protein copy number in \eqref{Reaction1} by $\widetilde{B}_2$, we have
\begin{equation}\label{accuracy}
\begin{aligned}
    \lvert B_{2}- \widetilde{B}_2\rvert\leq&\frac{v(u+v)^2}{2u^3(u+\delta)}
   \lVert D_1\rVert_{\infty}\\&
   +\frac{(u+v)^4}{2u^5\delta(u+\delta)}
   \lVert D_1\rVert_{\infty}^2
   \lVert D\rVert_{\infty}.
\end{aligned}
\end{equation}
According to \eqref{accuracy}, the absolute difference converges to zero at the rate of $O(u^{-2})$ when $u\rightarrow\infty$ and the other parameters remain fixed, as numerically verified in \autoref{fig:error_bound}. Notably, $u/\delta\gg1$ does not guarantee the validity of the burst approximation. 

In this study, we systematically analyze gene expression models under the burst approximation with multiple gene states and arbitrary burst size distributions. An analytical solution to the corresponding CME is established and exploited in two directions. On the one hand, we derive several inequalities for the PMF, the moment generating function, and the tail probability, using functional analysis. When the burst size is geometrically distributed, the probability distribution of the protein copy number is bounded by a scaled negative binomial distribution, and is light-tailed if the burst size parameter lies in $(0,1/2)$. The tail probability is estimated using the Chernoff method. On the other hand, we propose fast solvers for the steady-state distribution of the protein count in three special cases, namely, when the burst size is geometrically distributed, Poisson-distributed, or negligible beyond a certain threshold. Additionally, we show that our results are consistent with \cite{AnalyticalDistributionsStochastic2008}, and examine the validity of the burst approximation by estimating the approximation error in terms of low-order binomial moments. We conclude that, although \eqref{Reaction} is usually seen as a surrogate for full gene expression models, only the expectation of the protein number distribution is preserved, and higher-order moments generally differ. 

Several directions for future work naturally arise from this study.
First, the analytical solution to the CME may be further analyzed to estimate the mixing time of the corresponding Markov chain. Second, for the geometrically distributed burst size with parameter $\lambda\in[1/2,1)$, we have identified examples in which the binomial moment method fails to recover the PMF. How to compute the PMF in such cases remains open, and general conditions under which the binomial moment method is guaranteed to work also remain to be identified.
Third, models formulated via the CME implicitly assume the Markov property, whereas many physical systems are highly non-Markovian. This suggests that stochastic gene expression modeling should be extended to more general stochastic processes, such as those governed by the generalized chemical master equation \cite{ChemicalContinuousTime2017}.
Finally, in practice, the model validity is typically assessed from a top-down perspective. For example, the model parameters can be determined by fitting the computed protein number distribution to experimental data \cite{InferringKineticsStochastic2013,ApproximationInferenceMethods2017,InferringTranscriptionalBursting2023}. The development of appropriate statistical inference frameworks remains an important direction for future research.

\textit{Acknowledgments} -- We thank P. Xie and R. Gao for fruitful discussions. Y.L. was supported by Natural Science Foundation of China (125B10002). Y.Z. was supported by National Key R$\&$D Program of China (2024YFA1012401), the Science and Technology Commission of Shanghai Municipality (23JC1400501), and Natural Science Foundation of China (12241103).


\textit{Data availability} -- The Python code will be released in \url{https://github.com/yuntao2022} soon.

\bibliographystyle{siam}
\bibliography{Translation}
\end{document}